\documentclass[showpacs,showkeys,preprintnumbers,amsmath,amssymb]{revtex4}

\usepackage{graphicx}
\usepackage{dcolumn}
\usepackage{bm}
\begin{document}
\title{Neutrino nucleus cross sections for low energy neutrinos at SNS facilities}
\author{M. Sajjad Athar, Shakeb Ahmad }
\author{S.K.Singh}%
\affiliation{Physics Department, A. M. U., Aligarh-202 002, India.}
 \date{\today}

\begin{abstract}
  We calculate the neutrino nucleus cross sections for charged lepton production relevant for the experiments proposed with the stopped 
muon neutrinos using neutron spallation source facility. The calculations are done in local density approximation taking into account Pauli blocking, Fermi motion effects and renormalization of weak transition strengths in the nuclear medium. The effect of Coulomb distortion of the lepton produced in charge current reactions is taken into account by using the Fermi function as well as in a model where an effective momentum has been used for the lepton moving in the local Coulomb field of the final nucleus. The numerical results for the neutrino nucleus total cross sections averaged over Michel spectrum are presented for various nuclei.
\end{abstract}
\maketitle 
\section{ Introduction}
The study of neutrino nucleus cross sections at low energies is important in understanding various physical and astrophysical processes. The low energy neutrino beams $(E_\nu \le 52.8MeV)$ are generally obtained from the muons decaying at rest. The neutrino energy spectrum obtained from the muons decaying at rest is given by   
 \begin{equation}
\phi(E_{\nu_e})=\frac{12}{E_0^4}E_{\nu_e}^2(E_0-E_{\nu_e}),~~E_0=52.8MeV
\end{equation}
This is known as Michel spectrum and is shown in Fig.1.
\begin{figure}[h]
\includegraphics{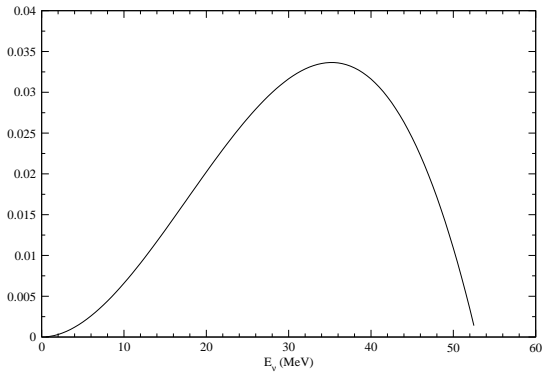}
\caption{Michel spectrum for $\nu_e$ obtained from muon decay at rest.}
\label{fg:Fig1}
\end{figure}\\
This neutrino energy spectrum and its energy range is similar to the energy spectrum and energy range of neutrinos coming from the core collapse supernova\cite{mclaughlin}. This similarity in the energy range and spectrum of the supernova neutrinos with the muons decaying at rest opens up the possibility of connecting the ground based neutrino nuclear experiments with the study of neutrino nuclear cross sections in supernova. Such a study will also be useful in understanding the r-process nucleosynthesis leading to the formation of heavy elements in the interstellar medium\cite{raffelt}.

The low energy neutrino beams can also be obtained at the Spallation Neutron Source(SNS) facilities where low energy neutrino nucleus reaction experiments can be performed\cite{conf}. At these facilities the accelerated protons hit a  nuclear target and produce $\pi^-$ and $\pi^+$ particles which give rise to muon and neutrinos as their decay product. There is a small flux of $\bar\nu_e$ as a result of $\mu^-$ decay as most of $\mu^-$ are captured by nuclei. On the other hand $\mu^+$ are not captured but are stopped in nuclei and decay at rest giving rise to substantial flux of $\nu_e$. These neutrinos can be used to study neutrino nuclear cross sections. Experiments to study the feasibility of making such neutrino-nucleus cross section measurements using many nuclear targets are proposed at Oak Ridge National Laboratory(ORNL) using stopped muon neutrino facility at the Spallation Neutron Source(SNS). Such a facility is expected to produce $\approx 10^{15}$ neutrinos per second of each flavor of $\nu_e$, $\nu_\mu$ and $\bar\nu_\mu$ with mono-energetic spectra for $\nu_\mu$  and continuous spectra for $\nu_e$ and $\bar\nu_\mu$. Out of these flavors of neutrinos, $\nu_e$ produces electrons which are observed to study the neutrino nuclear cross sections for charged current reactions.

The electron production from nuclei induced by low energy $\nu_e$ beams has been studied in past by many authors. A review of theoretical calculations and experimental results have been given by Donnelly and Peccei\cite{donnelly}, Kubodera et al.\cite{kubodera} and more recently by Kolbe et al.\cite{kolbe1}. It is proposed that these calculations are done for a large range of medium and heavy nuclei and attempts be made to study them experimentally. Such a list of nuclei where the study of neutrino nuclear cross section is of some interest has been recently prepared by Avignone and Efremenko\cite{avignone}.

In this paper we have studied neutrino nuclear cross sections induced by low energy $\nu_e$ obtained from muons decaying at rest for the nuclei proposed by Avignone and Efremenko\cite{avignone}. The calculations have been done in local density approximation taking into account Pauli blocking, Fermi motion effects and renormalization of weak transition strengths in the nuclear medium. The method has been successfully applied to study the various electromagnetic and weak processes in nuclei at low and intermediate energies\cite{singh}-\cite{athar}. The effect of Coulomb distortion of the lepton produced in charge current reactions is taken into account by using the Fermi function $F(Z,E_e)$, where Z is the atomic number and $E_e$ is the outgoing electron energy as well as in a Modified Momentum Approximation(MEMA), where the effect of Coulomb distortion is incorporated by modifying the momentum and energy of charged lepton in the Coulomb potential of the final nucleus\cite{kolbe1},\cite{engel}. 

In section-II, we describe the formalism. In section-III, we present the numerical results for the total cross section $\sigma(E) ~vs~$ the neutrino energy E and the flux averaged cross section and discuss the nuclear medium effects in some detail. A brief summary and conclusion of our results is given in section-IV.

\section{Formalism}
In this section we derive an expression for the total scattering cross section $\sigma(E_\nu)$ for the charged current(CC) reaction 
\begin{equation}
\nu_e +  ~_{Z_i}^{A}X  \rightarrow e^- +  ~_{Z_f}^{A}Y,
\end{equation}
where $Z_i(Z_f)$ is the charge of initial(final) nucleus. 

The basic $\nu_e$-neutron reaction taking place in $_{Z_i}^{A}X$ nucleus is
\begin{equation}
\nu_e(k) + n(p) \rightarrow e^-(k^\prime) + p(p^\prime),
\end{equation}
where k and p are the four momenta of the incoming neutrino and neutron and $k^\prime$ and $p^\prime$ are the four momenta of the outgoing electron and proton respectively.

The matrix element for the basic neutrino process on free nucleon (Eqn.3) is written as
\begin{equation}
T = \frac{G_F}{\sqrt{2}}\cos{\theta_c} ~l_{\mu} ~J^{\mu},
\end{equation}
where
\begin{eqnarray}
l_{\mu}&=&\bar{u}(k^\prime)\gamma_\mu(1-\gamma_5)u(k)\nonumber\\ 
J^\mu&=&\bar{u}(p^\prime)[F_{1}^V(q^2)\gamma^\mu \nonumber\\ 
&&+F_{2}^V(q^2)i{\sigma^{\mu\nu}}{\frac{q_\nu}{2M}} + F_{A}^V(q^2)\gamma^\mu\gamma_5]u(p).
\end{eqnarray}

The form factors $F_1^V(q^2)$, $F_2^V(q^2)$ and $F_A^V(q^2)$, where $q^2$ is the four momentum transfer square i.e. $(k-k^\prime)^2$, are isovector form factors and are written as
\[F_1^V(q^2)=F_1^p(q^2)-F_1^n(q^2), 
~F_2^V(q^2)=F_2^p(q^2)-F_2^n(q^2),\]
\[F_A^V(q^2)=F_A(q^2)\]
where
\begin{eqnarray}
 F^{p,n}_{1}(q^2)&=&\left[{1-\frac{q^2}{4M^2}}\right]^{-1}\left[{G^{p,n}_E(q^2)-\frac{q^2}{4M^2}G^{p,n}_M(q^2)}\right]~~\mbox\nonumber\\
 F^{p,n}_2(q^2)&=&\left[{1-\frac{q^2}{4M^2}}\right]^{-1}\left[{G^{p,n}_M(q^2)-G^{p,n}_E(q^2)}\right]\nonumber\\
\end{eqnarray}
\[G^p_E(q^2)=\frac{1}{(1-\frac{q^2}{M^2_v})^2}\]
\[G^p_M(q^2)=(1+\mu_p)G^p_E(q^2),~G^n_M(q^2)=\mu_n G^p_E(q^2);  \]
\[G^n_E(q^2)=(\frac{q^2}{4M^2})\mu_n G^p_E(q^2) \xi_n;~\xi_n=\frac{1}{1- \lambda_n\frac{q^2}{4M^2}}\]
\[\mu_p=1.792847, \mu_n=-1.913043, M_v=0.84GeV, ~\mbox{and}  ~\lambda_n=5.6.\]
The isovector axial vector form factor ${F_A}(Q^2)$ is given by
\[ {F_A}(Q^2)=\frac{F_A(0)}{({1-\frac{q^2}{M^2_A})}^2}\]
where $M_A=1.05GeV$; $F_A(0)$=-1.26

The double differential cross section  $\sigma_0(E_e,|\vec k^\prime|)$ for the basic reaction described in Eqn.(3) is then written as
\begin{equation}
\sigma_0(E_e,|\vec k^\prime|)={G_F}^2\cos^2{\theta_c}\frac{{|\vec k^\prime|}^2}{8\pi E_{\nu_e} E_e}\frac{M_n M_p}{E_n E_p}{\bar\Sigma}\Sigma|T|^2\delta[q_0+E_n-E_p]
\end{equation}
where ${\bar{\Sigma}}{\Sigma} {|T|^2}$ is the square modulus of the transition amplitude given by
\begin{eqnarray}
{\bar{\Sigma}}{\Sigma} {|T|^2}&=&L_{\mu\nu}J^{\mu\nu}~~with\mbox\nonumber\\
L_{\mu\nu}&=&8\left[k_\mu {k^\prime}_\nu + k_\nu {k^\prime}_\mu - k.k^\prime g_{\mu\nu} + i\epsilon_{\mu\nu\lambda\sigma}{k}^\lambda{k^\prime}^\sigma\right]~~and \mbox\nonumber\\
J^{\mu\nu}&=&\sum{J^\mu}^\dagger J^\nu
\end{eqnarray}

$q_0(q_0=E_{\nu_e} - E_e)$ is the energy transferred to the nucleon.

In a nucleus, the neutrino scatters from a neutron moving in the finite nucleus of neutron density $\rho_n(r)$, with a local occupation number $n_n({\bf{p}},{\bf{r}})$. In the local density approximation the scattering cross section is written as
\begin{equation}
\sigma(E_e,|\vec k^\prime|)=\int 2d{\bf r}d{\bf p}\frac{1}{(2\pi)^3}n_n({\bf p},{\bf r})\sigma_0(E_e,k^\prime)
\end{equation}
where $\sigma_0(E_e,|\vec k^\prime|)$ is given by Eqn.(7). The neutron energy $E_n$ and proton energy $E_p$ are replaced by $E_n(|\vec p|)$ and $E_p(|\vec{p}+\vec{q}|)$ where $\bf{p}$ is now the momentum of the target neutron inside the nucleus. In the nucleus the neutrons and protons are not free and their momenta are constrained to satisfy the Pauli principle, i.e., ${p_{n}}<{p_{F_{n}}}$ and ${p^\prime}_{p}(=|{\bf p}_{n}+{\bf q}|) > p_{F_{p}}$, where $p_{F_n}$ and $p_{F_p}$ are the local Fermi momenta of neutrons and protons at the interaction point in the nucleus and are given by $p_{F_n}=\left[3\pi^2\rho_n(r)\right]^\frac{1}{3}$ and $p_{F_p}=\left[3\pi^2\rho_p(r)\right]^\frac{1}{3}$, $\rho_n(r)$ and $\rho_p(r)$ are the neutron and proton nuclear densities. Furthermore, in nuclei the threshold value of the reaction i.e. the Q-value of the reaction has to be taken into account.

To incorporate these modifications, the $\delta$ function in Eqn.(7) i.e.  $\delta[q_0+E_n-E_p]$ is modified to $\delta[q_0+E_n(\vec{p})-E_p(\vec{p}+\vec{q})-Q]$ and the factor
\begin{equation}
\int \frac{d\bf{p}}{(2\pi)^3}{n_n(\bf{p},\bf{r})}\frac{M_n M_p}{E_n E_p}\delta[q_0+E_n-E_p]
\end{equation}
occurring in Eqn.(9) is replaced by $-(1/{\pi})$Im${{U_N}(q_0,\vec{q})}$, where ${{U_N}(q_0,\vec{q})}$ is the Lindhard function corresponding to the particle hole(ph) excitation shown in Fig.(2) and is given by
\begin{equation}
{U_N}(q_0,\vec{q}) = {\int \frac{d\bf{p}}{(2\pi)^3}\frac{M_nM_p}{E_nE_p}\frac{n_n(p)\left[1-n_p(\vec p + \vec q) \right]}{q_0+{E_n(p)}-{E_p(\vec p+\vec q)}+i\epsilon}}
\end{equation}
where $q_0$=$E_{\nu_e}-E_e-Q$.
 
\begin{figure}
\includegraphics{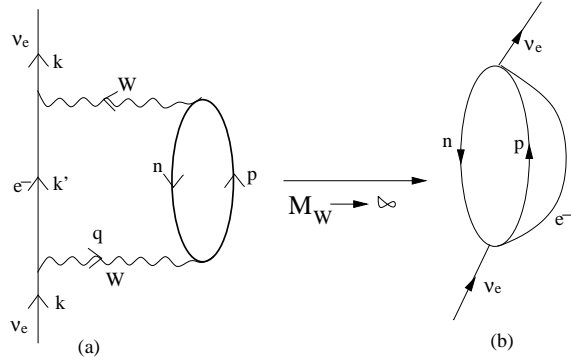}
\caption{Diagrammatic representation of the neutrino self-energy diagram corresponding to the ph-excitation leading to $\nu_e +n \rightarrow e^- + p$ in nuclei. In the large mass limit of the IVB(i.e.$M_W\rightarrow \infty$) the diagram 1(a) is reduced to 1(b) which is used to calculate $|T|^2$ in Eqn.(8).}
\label{fg:Fig1}
\end{figure}

The imaginary part of the Lindhard function is then derived to be: 
\begin{equation}
Im{U_N}(q_0, \vec{q}) = -\frac{1}{2\pi}\frac{M_p{M_n}}{|\vec{q}|}\left[E_{F_1}-A\right]~~~\mbox{with}
\end{equation}
$q^2<0$ ,$E_{F_2}-q_0<E_{F_1}$ and $\frac{-q_0+|\vec{q}|{\sqrt{1-\frac{4{M^2}}{q^2}}}}{2}<{E_{F_1}}$, 
where $E_{F_1}=\sqrt{p{_{F_n}}^2+{M_n}^2}$, $E_{F_2}=\sqrt{{p_{F_p}}^2+{M_p}^2}$ and \\
A = $Max\left[M_n,\hspace{2mm}E_{F_2}-q_0,\hspace{2mm}\frac{-q_0+|\vec{q}|\sqrt{1-\frac{4{M^2}}{q^2}}}{2}\right]$.\\

The threshold value, Q, for the neutrino reaction is generally taken to be the Q value corresponding to the lowest allowed Fermi or Gamow Teller transitions. However, in some cases the Q value corresponding to the ground state to ground state(gs-gs) transition is also taken\cite{haxton}.   

With inclusion of these nuclear effects the cross section $\sigma(E_\nu)$ is written as
\begin{eqnarray}
\sigma(E_\nu)=-\frac{2{G_F}^2\cos^2{\theta_c}}{\pi}\int^{r_{max}}_{r_{min}} r^2 dr \int^{{p_e}^{max}}_{{p_e}^{min}}{p_e}^2dp_e 
\int_{-1}^1d(cos\theta)\nonumber\\
\times\frac{1}{E_{\nu_e} E_e} L_{\mu\nu}J^{\mu\nu} Im{U_N}[E_{\nu_e} - E_e - Q, \vec{q}].
\end{eqnarray}
In the nucleus the strength of the electroweak coupling may change from their free nucleon values due to the presence of strongly interacting nucleons. Conservation of Vector Current (CVC) forbids any change in the charge coupling while magnetic and axial vector couplings are likely to change from their free nucleon values. These changes are calculated by considering the interaction of ph excitations in the nuclear medium in Random Phase Approximation (RPA) as shown in Fig.3. The diagram shown in Fig.3 simulates the effects of the strongly interacting nuclear medium at the weak vertex. The ph-ph interaction is shown by the wavy line in Fig.3 and is described by the $\pi$ and $\rho$ exchanges modulated by the effect of short range correlations.

The weak nucleon current described by Eq.(5) gives, in nonrelativistic limit, terms like $F_A \vec{\sigma}\tau_+$ and $i F_2 \frac{\vec{\sigma}\times \vec{q}}{2M}\tau_+$ which generate spin-isospin transitions in nuclei. While the term $i F_2 \frac{\vec{\sigma}\times \vec{q}}{2M}\tau_+$ couples to the transverse excitations, the term  $F_A \vec{\sigma}\tau_+$ couples to the transverse as well as longitudinal channels. These channels produce different RPA responses in the longitudinal and transverse channels when the diagrams of Fig.3 are summed over. This is illustrated by considering a term like $F_A\sigma^i$ in Eq.(5). One of the contributions of this term to the hadronic tensor $J^{ij}$ in the medium is proportional to $F^2_A \delta_{ij}Im U_N$ which is now split between the longitudinal and transverse components as 

\begin{figure}
\includegraphics{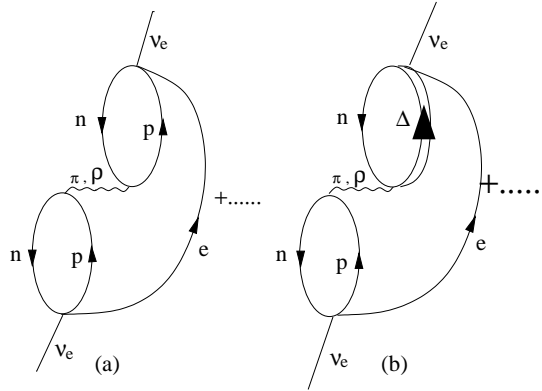}
\caption{ Many body Feynman diagrams (drawn in the limit $M_W\rightarrow \infty$) accounting for the medium polarization effects contributing to the process $\nu_e +n \rightarrow e^- + p$ transitions}
\label{fg:Fig1}
\end{figure}

\begin{equation}
F^2_A\delta_{ij}Im {U_N}\rightarrow F^2_A \left[{\bf{\hat{q_i}}{\hat{q_j}}}+(\delta_{ij}-{\bf{\hat{q_i}}{\hat{q_j}}})\right]Im {U_N}
\end{equation}
The RPA response of this term after summing the higher order diagrams like Fig.3 is modified and is given by $J^{ij}_{RPA}$:
\begin{equation}
J^{ij}\rightarrow J^{ij}_{RPA}= F^2_A{Im U_N}\left[\frac{{\bf{\hat{q_i}}{\hat{q_j}}}}{1-U_NV_l}+\frac{\delta_{ij}-{\bf{\hat{q_i}}{\hat{q_j}}}}{1-U_NV_t}\right]
\end{equation}
where $V_l$ and $V_t$ are the longitudinal and transverse part of the nucleon-nucleon potential calculated with $\pi$ and $\rho$ exchanges and are given by 
\begin{eqnarray}
V_l(q) = \frac{f^2}{m_\pi^2}\left[\frac{q^2}{-q^2+m_\pi^2}{\left(\frac{\Lambda_\pi^2-m_\pi^2}{\Lambda_\pi^2-q^2}\right)^2}+g^\prime\right],\nonumber\\
V_t(q) = \frac{f^2}{m_\pi^2}\left[\frac{q^2}{-q^2+m^2_\rho}{C_\rho}{\left(\frac{{\Lambda_\rho}^2-m^2_\rho}{{\Lambda_\rho}^2-q^2}\right)^2}+g^\prime\right]\end{eqnarray}
$\Lambda_\pi=1.3 GeV$, $C_\rho=2$, $\Lambda_\rho=2.5GeV$, $m_\pi$ and $m_\rho$ are the pion and $\rho$ masses, and $g^\prime$ is the Landau-Migdal parameter taken to be $0.7$ which has been used quite successfully to explain many electromagnetic and weak processes in nuclei\cite{mukh},\cite{gil}.

This modified tensor $J^{ij}_{RPA}$ when contracted with the leptonic tensor $L_{ij}$ gives the contribution of the $F^2_A$ term to the RPA response. The contribution of the time component of the hadronic tensors like $J^{0i}$ and $J^{00}$ are of higher order in $\left(\frac{q}{M}\right)^2$ and are not important at low energies considered in this paper.

The effect of the $\Delta$ degrees of freedom in the nuclear medium is included in the calculation of the RPA response by considering the effect of ph-$\Delta$h and $\Delta$h-$\Delta$h excitations as shown in Fig.3(b). This is done by replacing $U_N$ by $U_N=U_N+U_\Delta$, where $U_\Delta$ is the Lindhard function for $\Delta$h excitation in the medium and the expressions for $U_N$ and $U_\Delta$ are taken from\cite{oset1}. The different couplings of $N$ and $\Delta$ are incorporated in $U_N$ and $U_\Delta$ and then the same interaction strengths $V_l$ and $V_t$ are used to calculate the RPA response. This is discussed in some detail in Ref\cite{singh} and more recently in Ref.\cite{nieves} by Nieves et al.

One of the important aspects of charge current neutrino interactions is the treatment of Coulomb distortion of the produced lepton in the Coulomb field of the final nucleus. At low energies of the electron relevant to $\beta$ decays in nuclei the Coulomb distortion of electron in the nuclear field is taken into account by multiplying the momentum distribution of the electron in Eqn.(13) by a Fermi function $F(Z,E_e)$, where $F(Z,E_e)$ is given by\cite{behrens}:
\[F(Z,E_e)=\left[1-\frac{2}{3}(1-\gamma_0)\right]^{-1}f(Z, E_e),\]

where \[f(Z, E_e)=2(1+\gamma_0))(2p_eR)^{-2(1-\gamma_0)}\frac{|\Gamma(\gamma_0+i\eta)|^2}{(\Gamma(2\gamma_0+1))^2}.\]  
Here R is the nuclear radius and $\gamma_0=\sqrt{1-(\alpha Z)^2}$, $\eta=\frac{\alpha Z c}{v}$. This approximation works quite well at low energies, but it is not appropriate at higher energies, specially for high Z nuclei\cite{kolbe1},\cite{engel}. Therefore, at higher electron energies a different approach is needed to describe the Coulomb distortion effect of the electron. For this purpose, we apply the methods of electron scattering where various approximations have been used to take into account the Coulomb distortion effects of the initial and final electron\cite{gco}-\cite{guisti}. One of them is the Modified Effective Momentum Approximation(MEMA) in which the electron momentum and energy are modified by taking into account the Coulomb energy. We have used this approach in the case of charged current quasielastic neutrino scattering and the energy and momentum of the electron present in the final state is modified in the Coulomb field of the final nucleus\cite{mukh},\cite{argon}-\cite{athar}. In the local density approximation, the effective energy of the electron in the Coulomb field of the final nucleus is given by:
\[ E_{eff} = E_e + V_c(r), \]
where 
\begin{equation}
V_c(r)=Z_f\alpha4\pi(\frac{1}{r}\int_0^r\frac{\rho_p(r^\prime)}{Z_f}{r^\prime}^2dr^\prime + \int_r^\infty\frac{\rho_p(r^\prime)}{Z_f}{r^\prime}dr^\prime)
\end{equation}

Thus, in presence of nuclear medium effects the total cross section $\sigma(E_\nu)$, with the inclusion of Coulomb distortion effects taken into account by Fermi function(MEMA), is written as
 \begin{eqnarray}
\sigma^{FF(MEMA)}(E_\nu)=-\frac{2{G_F}^2\cos^2{\theta_c}}{\pi}\int^{r_{max}}_{r_{min}} r^2 dr \int^{{p_e}^{max}}_{{p_e}^{min}}{p_e}^2dp_e 
\int_{-1}^1d(cos\theta)\nonumber\\
\times\frac{1}{E_{\nu_e} E_e} L_{\mu\nu}{J^{\mu\nu}_{RPA}} Im{U_N}^{FF(MEMA)}.
\end{eqnarray}
where 
\begin{eqnarray}
Im{U_N^{FF}}&=&F(Z,E_e)Im{U_N}[E_{\nu_e} - E_e - Q, \vec{q}]~~ and \mbox\nonumber\\
Im{U_N^{MEMA}}&=&Im{U_N}[E_{\nu_e} - E_e - Q-V_c(r), \vec{q}]
\end{eqnarray}
\section{Results}
In this section we present the results for the total cross section $\sigma(E)$ as a function of energy and the flux averaged cross section $<\sigma>$ for various nuclei which have been presently proposed to be studied at SNS facilities using neutrinos from stopped muon decays\cite{avignone}. For the numerical calculation of the cross section, we have classified the nuclei in three groups according to their nuclear densities used in this calculation and have been presented in Tables I-III. 
 
In Table-I, we present the nuclear density parameters for $^{12}{C}$, $^{14}{N}$ and $^{16}{O}$ nuclei using a 2-parameter harmonic oscillator(H.O.) density given by
\begin{equation}
\rho(r)=\rho_0(1+\alpha(\frac{r}{a})^2)exp(-(\frac{r}{a})^2),
\end{equation}

 In Table-II, we present the nuclear density parameters for $^{19}{F}$, $^{23}{Na}$, $^{27}{Al}$, $^{28}{Si}$, $^{31}{P}$, $^{37}{Cl}$, $^{40}{Ar}$, $^{51}{V}$, $^{52}{Cr}$, $^{55}{Mn}$, $^{56}{Fe}$, $^{59}{Co}$, $^{71}{Ga}$, $^{89}{Y}$, $^{93}{Nb}$, $^{98}{Mo}$, $^{115}{In}$, $^{127}{I}$, $^{139}{La}$, $^{181}{Ta}$, $^{208}{Pb}$ and $^{209}{Bi}$ nuclei using a 2-parameter Fermi density (2pF) given by
\begin{equation}
\rho(r)=\frac{\rho_0}{(1+exp((r-\alpha)/a))}
\end{equation}
In Table-III,  we present the nuclear density parameters for $^{32}{S}$, $^{39}{K}$ and $^{40}{Ca}$ nuclei using a three parameter Fermi(3pF) density given by
\begin{equation}
\rho(r)=\frac{\rho_0(1+w\frac{r^2}{\alpha^2})}{(1+exp((r-\alpha)/a))},
\end{equation}
 The parameters have been taken from de Vries et al.\cite{vries} except for $^{115}{In}$ and $^{127}{I}$ which have been taken from Ref.\cite{kosmas}. The Q values presented in these tables correspond to the lowest allowed Fermi or Gamow-Teller transitions for the above mentioned nuclei except for the case of $^{40}Ca$ and $^{98}Mo$ for which the Q value corresponding to the ground state to ground state transitions have been taken\cite{haxton},\cite{table}. 

\subsection {Nuclear Medium Effects}
When the reaction $\nu_e + n \rightarrow e^- + p$ takes place in the nucleus, the first consideration is the Q value which inhibits the reaction in the nucleus. This inhibition is quite substantial in the low energy region considered here for the nuclei like $^{12}C$, $^{16}O$, $^{18}Si$, $^{32}S$ and $^{40}Ca$ for which the Q values are rather large (Q$\approx$13-18MeV). In addition to this, the effect of Pauli blocking which is taken into account through the imaginary part of the Lindhard function is to further reduce the cross section. Finally the renormalisation of weak coupling constants which is generated in our model through RPA correlations and is taken into account by calculating the cross section with the modified hadronic tensor ${J^{\mu\nu}_{RPA}}$ also reduces the cross sections. In Fig.4, we have shown the reduction due to these effects separately for some representative nuclei like $^{12}C$, $^{56}Fe$, $^{127}I$ and $^{208}Pb$ in various mass ranges.

We see that at low energies the major suppression in the cross section comes due to the consideration of Q-values and Pauli blocking in the nuclear medium. The reduction in the cross section $\sigma(E)$ due to these effects decreases with the increase of energy. For example at $E_\nu=50MeV$, this suppression is $\approx$ $93\%$ for $^{12}C$ and $\approx$ $75-77\%$ for other nuclei like $^{56}Fe$, $^{127}I$ and $^{208}Pb$ respectively (compare the solid lines with the dashed line in Fig.4). This suppression reduces to $40-45\%$ in all these nuclei at $E_\nu=200MeV$(not shown in Fig.4). 

In addition to the Pauli blocking, the consideration of RPA correlation in the nuclear medium gives rise to further reduction which increases with the mass number and decreases with the increase in energy (compare the dashed line with the dotted line in Fig.4). For example at $E_\nu=50MeV$ the RPA correlations give a further reduction of $50\%$ for $^{12}C$, $60\%$ for $^{56}Fe$ and around $70\%$ for $^{127}I$ and $^{208}Pb$. As the energy increases it becomes smaller and at $E_\nu=200MeV$ the reduction is $35\%$ for $^{12}C$, $40\%$ for $^{56}Fe$ and around $50\%$ for $^{127}I$ and $^{208}Pb$(not shown here). It should be noted that $40-60\%$ reduction due to the medium polarisation effects calculated through the RPA correlations in our model is similar to using $\frac{g_{eff}}{g_A}=0.7$ in some shell model calculations\cite{kolbe2}-\cite{suzuki}.

\subsection {Effects of Coulomb distortion}
The effect of Coulomb distortion is calculated using Fermi function $F(Z,E_e)$ as well as with the modified effective momentum approximation(MEMA). The results for some representative nuclei like $^{12}C$, $^{56}Fe$, $^{127}I$ and $^{208}Pb$ in various mass range are shown in Fig.5. The general effect of the Coulomb distortion of the electron is to increase the cross section which depends upon the incident energy and the charge of the final nucleus.

For a fixed Z, this increase in the cross section decreases with the increase in energy while for a fixed energy the inclusion of Coulomb distortion increases with the charge Z. For example for $^{12}C$ this is $15\%$ at $E_\nu=50MeV$ which becomes $10\%$ at $E_\nu=200MeV$. For high Z nuclei the Coulomb effect is very large and results in manifold increase in the cross sections. This can be seen by comparing the cross section without Coulomb effect shown by dotted lines and the cross sections with Coulomb effects using the Fermi function $F(Z,E_e)$ shown by dashed lines. For example in the case of $^{56}Fe$ nucleus the increase due to Coulomb distortion is $83\%$ at $E_\nu=50MeV$ which becomes $75\%$ at $E_\nu=200MeV$. However, as discussed in Section-II, the use of Fermi function to calculate the Coulomb distortion effects overestimates the cross sections and is not appropriate at higher electron energies. Therefore, we use the modified effective momentum approximation(MEMA) and present the results for $\sigma(E)$ in Fig.5 with solid lines. It is seen that for low Z nuclei, like $^{12}C$. the results for the cross sections in the two approximations are qualititative similar but the MEMA gives slightly higher cross sections for $E_\nu<60MeV$. As the Z increases the cross sections with MEMA is higher than Fermi function at lower energies and becomes smaller than the cross sections obtained with the Fermi function as the energy becomes large. This energy dependence of the cross section is explicitly shown for higher Z nuclei like $^{56}Fe$, $^{127}I$ and $^{208}Pb$ in Fig.5. 

We show the results of the energy dependence of the cross sections for all nuclei listed in tables I-III, in Figs.6-9, where dashed lines show the cross sections without RPA correlations and dotted lines show the cross sections with RPA correlations using Fermi function for the Coulomb distortion. The solid lines show the cross section with RPA correlations where the Coulomb distortion effects are calculated with MEMA. In these figures a comparison of dashed lines and dotted lines shows the effect of RPA correlations while a comparison of dotted lines and solid lines shows the effect of Coulomb distortion calculated using Fermi function and MEMA.

\subsection {Flux averaged cross sections}
We calculate the flux averaged cross section $<\sigma>$ defined as 
\begin{equation}
<\sigma>=\int\phi(E_\nu)\sigma(E_\nu)dE_\nu
\end{equation}
 where $\phi_(E_\nu)$ is given by Eqn.1. The results for $<\sigma>$ are presented in Table-IV, where we show by $<\sigma>^{RPA}_{NC}$ the flux averaged cross sections with nuclear medium effects without any Coulomb distortion effects. When Coulomb distortion effects are taken into account the cross sections without RPA correlations are shown by $<\sigma>^N_{C}$ and the results with RPA correlations are shown by $<\sigma>^{RPA}_{C}$. We evaluate $<\sigma>^{N}_{C}$ and $<\sigma>^{RPA}_{C}$ in a hybrid model where at lower energies $\sigma(E_\nu)$ calculated with the Fermi function  and at higher energies $\sigma(E_\nu)$ calculated with MEMA is used  to perform the flux averaging in Eqn.23. For low mass nuclei like $^{12}{C}$, $^{16}{O}$, etc. the flux averaged cross section has been evaluated with the cross section $\sigma(E)$ calculated with the Fermi function. Thus, in hybrid model, it is the lower value of the cross section which is used for calculating $<\sigma>$. We see from this table that the effect of the Coulomb distortion is to increase the cross section and this increase is quite large for high $Z$ nuclei like $^{56}{Fe}$, $^{208}{Pb}$, etc. In case of $^{12}C$, it is small but plays an important role in explaining the experimental result(Compare column 1 and column 3 in table IV). A comparison of column 2 and column 3 in this table shows the strong reduction due to RPA correlations which increases with mass number.

In Table-V, we compare our results with the results of some other calculations. In this energy region of the neutrinos, there are many theoretical calculations done for the inclusive neutrino reactions in $^{12}C$\cite{mukh}-\cite{nieves}, \cite{kolbe}-\cite{carbon}, while there are few calculations for $^{16}O$\cite{auer}, $^{56}Fe$\cite{kolbe2},\cite{mintz1} and $^{208}Pb$\cite{kolbe2},\cite{suzuki}-\cite{volpe1} and some other nuclei\cite{kosmas}. Some calculations are similar to the calculations presented in this paper\cite{kosmas}-\cite{nieves} while others make use of Shell Model\cite{volpe}-\cite{suzuki}, random phase approximation(RPA) with pairing correlations\cite{kolbe2}-\cite{volpe},\cite{kolbe}-\cite{volpe1} and elementary particle approach\cite{mintz}-\cite{mintz1}. We see that for $^{12}C$ and $^{56}Fe$ our results are in fair agreement with the experimental results and other theoretical calculations. For $^{208}Pb$ nucleus our results for $<\sigma>$ is comparatively smaller than the results of  Refs.\cite{kolbe2}, \cite{suzuki} and \cite{volpe1}. This is mainly due to the different approaches of taking into account the nuclear effects. However, among the different calculations of the inclusive cross section $<\sigma>$ in $^{208}Pb$, the results do not agree among themselves\cite{kolbe2},\cite{suzuki}-\cite{volpe1}. Therefore, more work is needed for calculating the cross section in $^{208}Pb$ at low energies.

\section{Summary and Conclusions}
We have studied the charged current $\nu_e$ reactions on various nuclei which are of present interest. The cross section calculations are performed in a local density approximation taking into account the Pauli blocking, Fermi motion and the renormalization of weak transition strengths in the nuclear medium. The effect of Coulomb distortion for the charged lepton while coming out of the nucleus is taken into account by using the Fermi function as well as the modified momentum approximation(MEMA). The cross sections are then averaged over the $\nu_e$ spectra obtained from the muons decay at rest where the maximum energy of neutrinos is 52.8MeV. 

We find that 

1. The role of nuclear effects like Q value, Pauli blocking and Fermi motion is to reduce the cross sections. For a given Z, this reduction becomes smaller with the increase in energy. For example, at $E_\nu=50MeV$, this suppression is $\approx$ $93\%$ for $^{12}C$ and $\approx$ $75-77\%$ for other nuclei like $^{56}Fe$, $^{127}I$, etc. This suppression reduces to $40-45\%$ in all these nuclei at $E_\nu=200MeV$.

2. There is a further reduction of the cross section due to the renormalization of weak transition strengths in the nuclear medium. For a given Z, this reduction becomes smaller with the increase in neutrino energy, while for a given neutrino energy $E_\nu$, this reduction increases with Z. For example at $E_\nu=50MeV$ the RPA correlations give a further reduction of $50\%$ for $^{12}C$, $60\%$ for $^{56}Fe$ and around $70\%$ for $^{127}I$ and $^{208}Pb$ and at $E_\nu=200MeV$ this reduction is $35\%$ for $^{12}C$, $40\%$ for $^{56}Fe$ and around $50\%$ for $^{127}I$ and $^{208}Pb$.

3. The two methods of treating the Coulomb distortion give similar results for low energy neutrinos in the case of low mass nuclei. For intermediate and heavy mass nuclei the cross sections with Fermi function are smaller than the cross sections with MEMA upto certain energy $E_{\nu_e}$ after which the the cross sections calculated with Fermi function become larger. At the energy $E_{\nu_e}$ where this cross over takes place changes with nuclei. For example it is around 40MeV for nuclei like $^{56}Fe$ in the intermediate mass range and around 18MeV for nuclei in the heavier mass range like $^{208}Pb$.

4. The total cross sections averaged over the neutrino spectrum obtained from the muons decaying at rest is presented for all nuclei considered here. The results for  $^{12}C$, $^{16}O$, $^{56}Fe$ and $^{208}Pb$ nuclei are compared with the available experimental results as well as different theoretical calculations. New results have been presented for many other nuclei. 
   
 The numerical results presented in this paper can be a very useful benchmark for neutrino nucleus cross section measurements being proposed at SNS facilities using various nuclei as nuclear targets. 
\acknowledgements
This work was financially supported by the Department of Science and Technology, Govt. of India under grant number DST SP/S2/K-07/2000. One of the authors(SA) would like to thank CSIR for the financial support. 
\begin{table}[h]
\caption{Q-value and the density parameters used in the numerical evaluation of the cross sections. $\alpha$ and $a$ are the parameters for HO density used in Eqn.(20).} 
\begin{ruledtabular}
\begin{tabular}{cccc}
Nucleus & Q(transition)&$\alpha$&a\\ \vspace{1mm}
 & (MeV)&&(fm)\\ \hline \\ \vspace{1mm}
$^{12}{C}$&17.84 ($0^+ \rightarrow 1^+$)&1.69&1.07  \\ \vspace{1mm}
$^{14}{N}$&5.60  ($1^+ \rightarrow 0^+$)&1.76&1.23  \\ \vspace{1mm}
$^{16}{O}$&19.76 ($0^+ \rightarrow 1^+$)&1.80&1.52 
\end{tabular}
\end{ruledtabular}
\end{table}

\begin{table}
\caption{Q-value and the density parameters used in the numerical evaluation of the cross sections. $\alpha$ and $a$ are the parameters used in Eqn.(21) for the 2pF density.} 
\begin{ruledtabular}
\begin{tabular}{cccccccc}
Nucleus & Q(transition)&$\alpha$&a&Nucleus & Q(transition)&$\alpha$&a\\ \vspace{1mm}
 & (MeV)&(fm)&(fm)&& (MeV)&(fm)&(fm)
\\ \hline\\ \vspace{1mm}
$^{19}{F}$&3.75 $({\frac{1}{2}}^+ \rightarrow {\frac{1}{2}}^+)$&2.59&0.56 &$^{59}{Co}$&1.92 $({\frac{7}{2}}^- \rightarrow {\frac{5}{2}}^-)$&4.08&0.57 \\ \vspace{1mm}
$^{23}{Na}$&4.56 $({\frac{3}{2}}^+ \rightarrow {\frac{3}{2}}^+)$&2.81&0.54&$^{71}{Ga}$&0.74 $({\frac{3}{2}}^- \rightarrow {\frac{1}{2}}^-)$&4.44&0.58   \\ \vspace{1mm}
$^{27}{Al}$&5.32 $({\frac{5}{2}}^+ \rightarrow {\frac{5}{2}}^+)$&3.07&0.52&$^{89}{Y}$&4.44 $({\frac{1}{2}}^- \rightarrow {\frac{3}{2}}^-)$&4.76&0.57 \\ \vspace{1mm}
$^{28}{Si}$&14.80 ($0^+ \rightarrow 0^+$)&2.93&0.57&$^{93}{Nb}$&0.91 $({\frac{9}{2}}^+ \rightarrow {\frac{7}{2}}^+)$&4.87&0.57 \\ \vspace{1mm}
$^{31}{P}$&5.91 $({\frac{1}{2}}^+ \rightarrow {\frac{1}{2}}^+)$&3.21&0.56&$^{98}{Mo}$&2.20($0^+ \rightarrow 6^+$)&5.10 &0.57 \\ \vspace{1mm}
$^{37}{Cl}$&1.32 $({\frac{3}{2}}^+ \rightarrow {\frac{3}{2}}^+)$&3.53&0.52&$^{115}{In}$&0.008 $({\frac{9}{2}}^+ \rightarrow {\frac{7}{2}}^+)$&5.36&0.56 \\ \vspace{1mm}
$^{40}{Ar}$&4.3 ($0^+ \rightarrow 1^+$)&3.39&0.61&$^{127}{I}$&1.42 $({\frac{5}{2}}^+ \rightarrow {\frac{3}{2}}^+)$&5.40&0.55 \\ \vspace{1mm}
$^{51}{V}$&1.26 $({\frac{7}{2}}^- \rightarrow {\frac{7}{2}}^-)$&3.94&0.50 &$^{139}{La}$&0.78 $({\frac{7}{2}}^+ \rightarrow {\frac{9}{2}}^+)$&5.71 &0.53 \\ \vspace{1mm}
$^{52}{Cr}$&5.22 ($0^+ \rightarrow 1^+$)&4.01&0.50 &$^{181}{Ta}$&0.70 $({\frac{7}{2}}^- \rightarrow {\frac{5}{2}}^-)$&6.38&0.64\\ \vspace{1mm}
$^{55}{Mn}$&0.74 $({\frac{5}{2}}^- \rightarrow {\frac{3}{2}}^-)$&3.89 &0.57&$^{208}{Pb}$&5.20 $(0^+ \rightarrow 1^+)$&6.62&0.50\\ \vspace{1mm}
$^{56}{Fe}$&6.82 ($0^+ \rightarrow 1^+$)&3.97&0.59&$^{209}{Bi}$&3.80 $({\frac{9}{2}}^- \rightarrow {\frac{7}{2}}^-)$&6.75&0.47\\  

\end{tabular}
\end{ruledtabular}
\end{table}

\begin{table}
\caption{Q-value and the density parameters used in the numerical evaluation of the cross sections. $\alpha$, $a$ and $w$ are the parameters used in Eqn.(22) for the 3pF density.} 
\begin{ruledtabular}
\begin{tabular}{ccccc}
Nucleus & Q(transition)&$\alpha$&a&w\\ \vspace{1mm}
& (MeV)&(fm)&(fm)&\\ \hline\\ \vspace{1mm}
$^{32}{S}$&13.20 $(0^+ \rightarrow 1^+)$&3.50&0.63&{-0.25}\\ \vspace{1mm}
$^{39}{K}$&7.04 $({\frac{3}{2}}^+ \rightarrow {\frac{3}{2}}^+)$&3.74&0.58&-0.201\\ \vspace{1mm}
$^{40}{Ca}$&14.80($0^+\rightarrow 4^-$)&3.67&0.58&-0.1017\\ 
\end{tabular}
\end{ruledtabular}
\end{table}

\begin{figure*}
\includegraphics{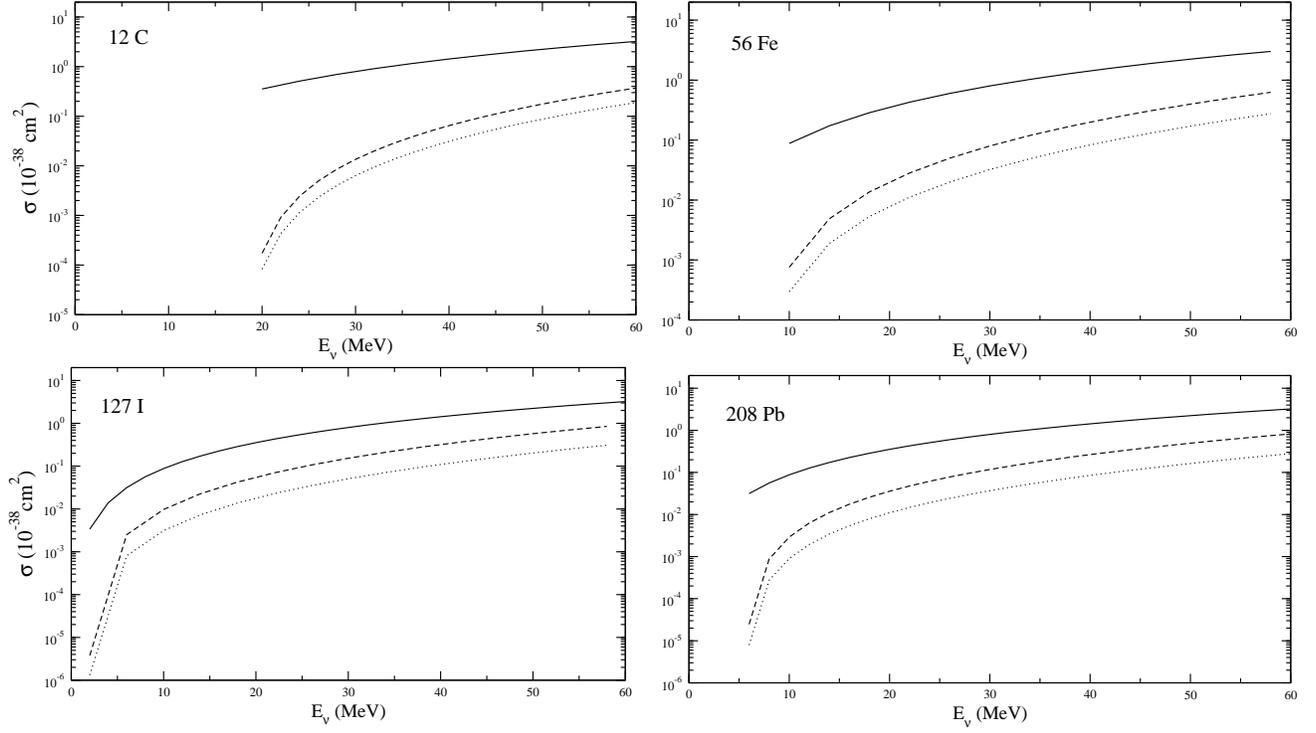}
\caption{Quasielastic cross section $\sigma(E_\nu)\sim E_\nu$ for neutrino reaction in some nuclei. The dashed (dotted) lines represent neutrino nucleus cross section without (with) RPA correlations to be compared with the free nucleon cross section (solid lines).}
\end{figure*}
\begin{figure*}
\includegraphics{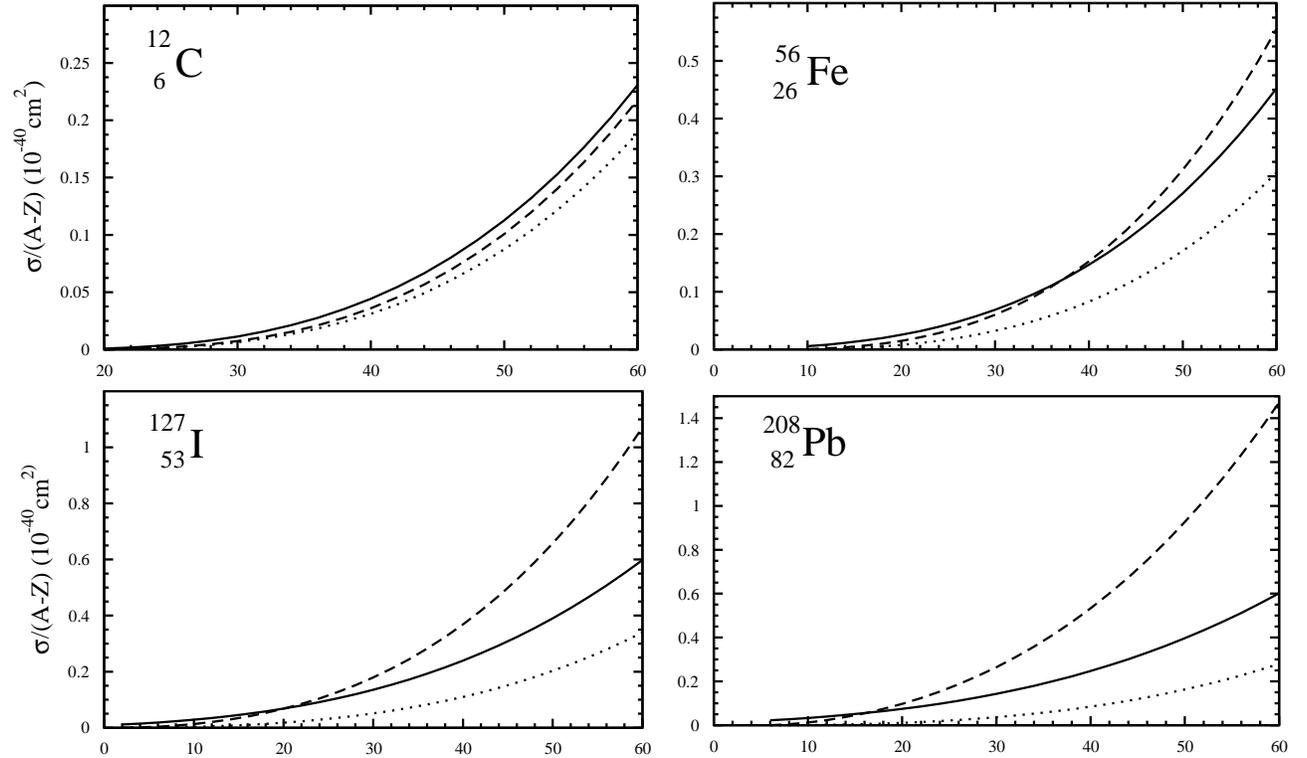}
\caption{Quasielastic cross section $\sigma(E_\nu)\sim E_\nu$ for neutrino reaction in some nuclei. The dashed (solid) lines show the cross section calculated with RPA correlations and Coulomb effects using Fermi function (MEMA) to be compared with the cross section without Coulomb effects (dotted lines).} 
\end{figure*}
\begin{figure*}
\includegraphics{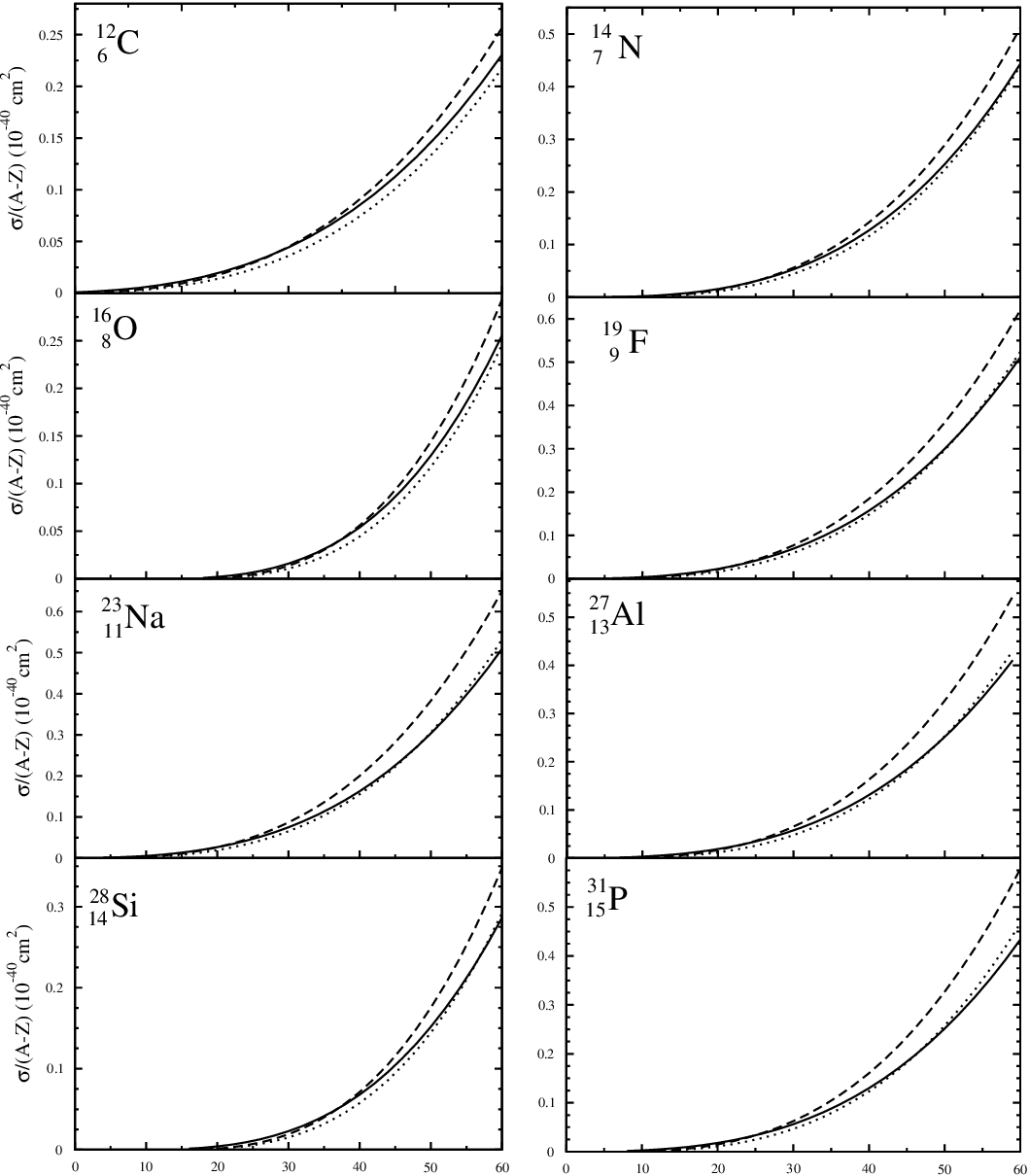}
\caption{Quasielastic cross section $\sigma(E_\nu)\sim E_\nu$ for neutrino reaction in various nuclei. The dotted (solid) lines show the cross section calculated with RPA correlations and Coulomb effect using fermi function (MEMA) to be compared with the cross section calculated with Coulomb effect using Fermi function and without RPA correlations (dashed lines-shown here after multiplying by 0.6).} 
\end{figure*}
\begin{figure*}
\includegraphics{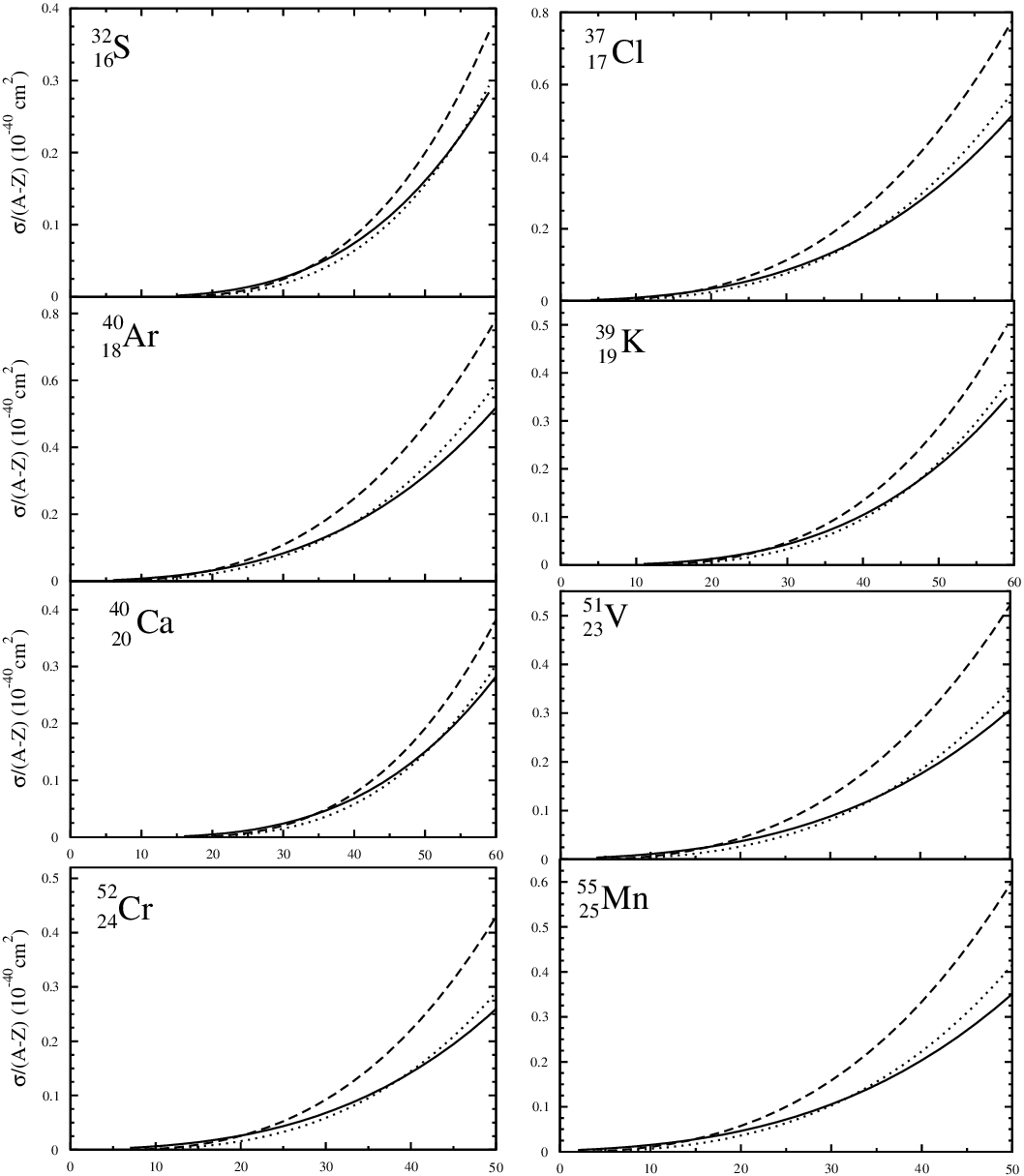}
\caption{Quasielastic cross section $\sigma(E_\nu)\sim E_\nu$ for neutrino reaction in various nuclei. The dotted (solid) lines show the cross section calculated with RPA correlations and Coulomb effect using Fermi function (MEMA) to be compared with the cross section calculated with Coulomb effect using Fermi function and without RPA correlations (dashed lines-shown here after multiplying by 0.6).} 
\end{figure*}
\begin{figure*}
\includegraphics{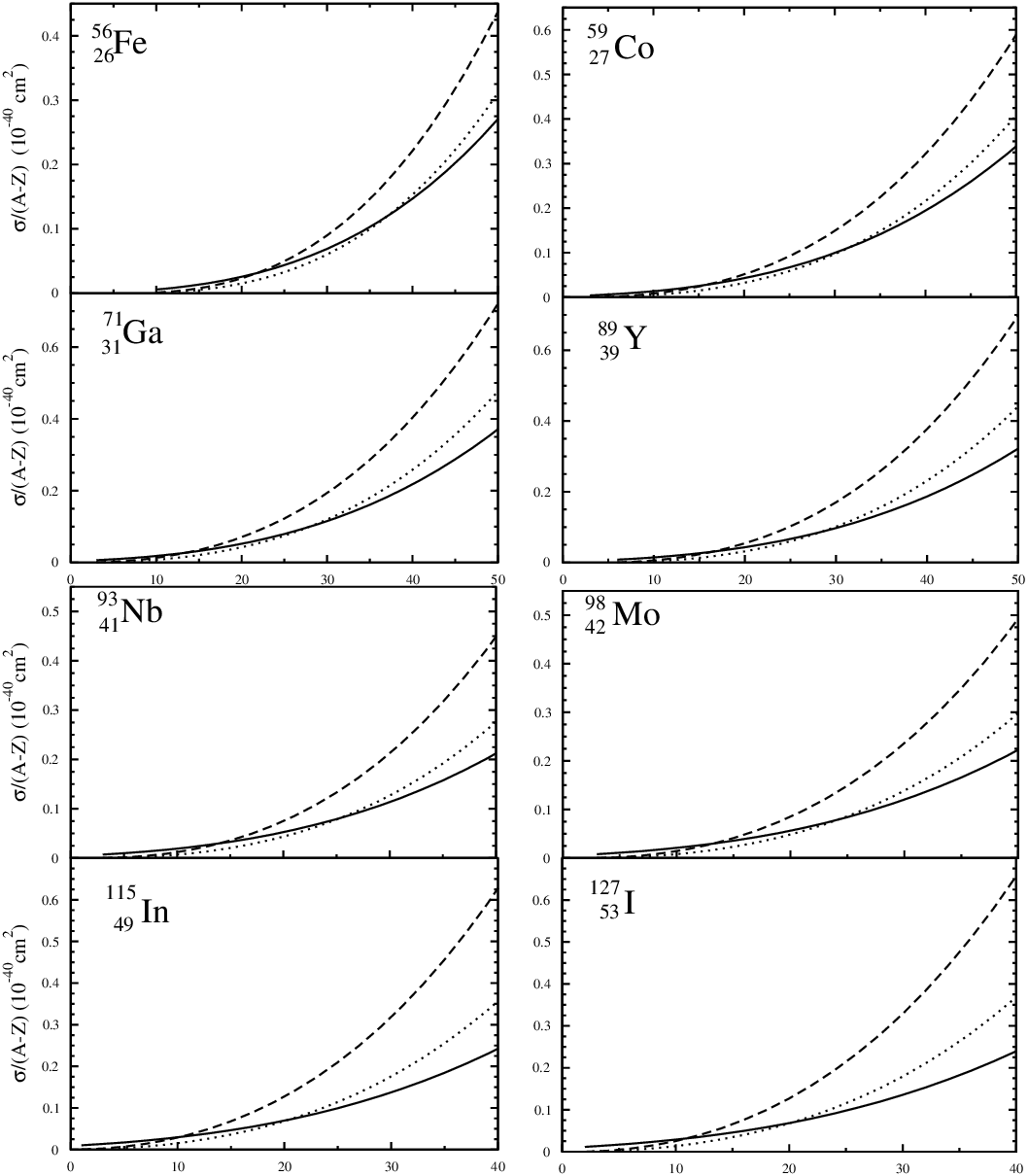}
\caption{Quasielastic cross section $\sigma(E_\nu)\sim E_\nu$ for neutrino reaction in various nuclei. The dotted (solid) lines show the cross section calculated with RPA correlations and Coulomb effect using Fermi function (MEMA) to be compared with the cross section calculated with Coulomb effect using Fermi function and without RPA correlations (dashed lines-shown here after multiplying by 0.6).} 
\end{figure*}
\begin{figure*}
\includegraphics{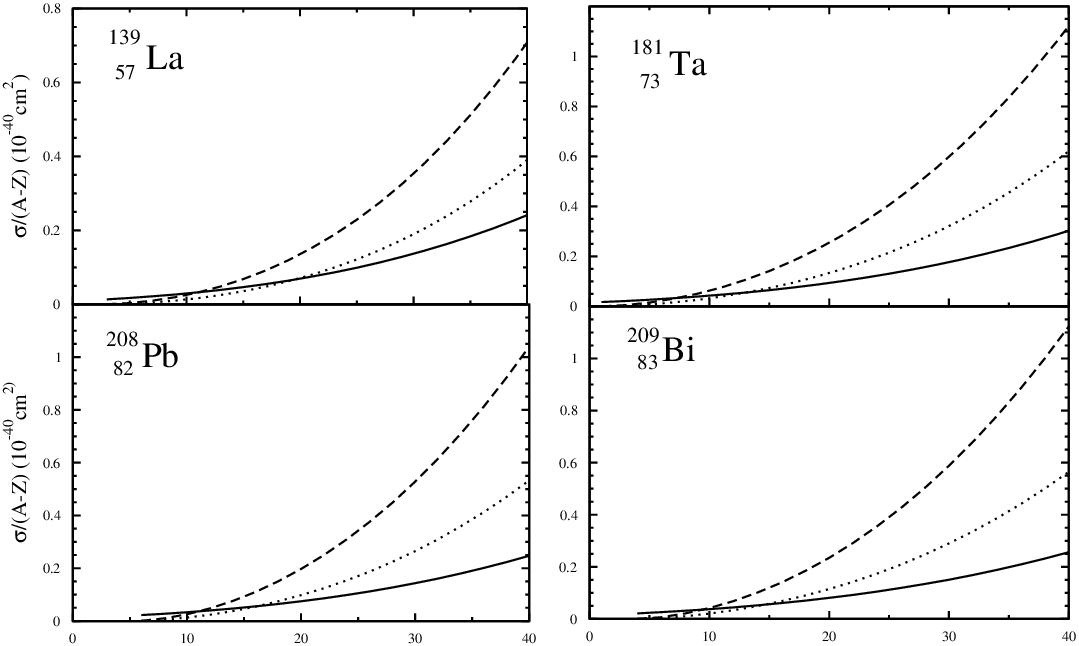}
\caption{Quasielastic cross section $\sigma(E_\nu)\sim E_\nu$ for neutrino reaction in various nuclei. The dotted (solid) lines show the cross section calculated with RPA correlations and Coulomb effect using Fermi function (MEMA) to be compared with the cross section calculated with Coulomb effect using Fermi function and without RPA correlations (dashed lines-shown here after multiplying by 0.6).} 
\end{figure*}

\begin{table}[h]
\caption{Total cross section $<\sigma>$(in $10^{-42}cm^2$) averaged over the Michel spectrum. $<\sigma>^{RPA}_{NC}$ is the averaged cross section calculated with RPA correlations without the Coulomb effect, $<\sigma>^{RPA}_C$ is the averaged cross section calculated with RPA correlations with Coulomb effect and $<\sigma>^{N}_{C}$ is the averaged cross section calculated without RPA correlations with the Coulomb effect.} 
\begin{ruledtabular}
\begin{tabular}{cccccccc}
Nucleus & $<\sigma>^{RPA}_{NC}$ & $<\sigma>^{N}_{C}$ &$<\sigma>^{RPA}_{C}$&Nucleus & $<\sigma>^{RPA}_{NC}$ & $<\sigma>^{N}_{C}$&$<\sigma>^{RPA}_{C}$
\\ \hline \\\vspace{2mm}
$^{12}_{6}{C}$&   11.80  &28.10&13.60    &$^{52}_{24}{Cr}$&  157.10& 643.18&254.70       \\ \vspace{2mm}
$^{14}_{7}{N}$&   44.60  &107.91&52.65    &$^{55}_{25}{Mn}$&  250.40& 1144.52&412.40       \\ \vspace{2mm}
$^{16}_{8}{O}$&   12.00  &30.43&14.55    &$^{56}_{26}{Fe}$&  161.40& 685.25&277.00       \\ \vspace{2mm}
$^{19}_{9}{F}$&   77.80  &201.07&96.35    &$^{59}_{27}{Co}$&  244.70& 1148.62&418.70       \\ \vspace{2mm}
$^{23}_{11}{Na}$& 94.70  &263.50&120.50   &$^{71}_{31}{Ga}$&  335.70& 1762.91&596.00       \\ \vspace{2mm}
$^{27}_{13}{Al}$& 82.17  &247.13&111.62   &$^{89}_{39}{Y}$&   307.10& 1912.11&633.50       \\ \vspace{2mm}
$^{28}_{14}{Si}$& 33.17  &95.18&46.25    &$^{93}_{41}{Nb}$ & 370.00& 2332.71&760.10       \\ \vspace{2mm}
$^{31}_{15}{P}$&  89.34  &276.60&127.10   &$^{98}_{42}{Mo}$&  417.60& 2661.25&854.00       \\ \vspace{2mm}
$^{32}_{16}{S}$&  44.10  &141.38&64.40    &$^{115}_{49}{In}$& 516.60& 3836.37&1133.00      \\ \vspace{2mm}
$^{37}_{17}{Cl}$& 154.70 &531.18&216.00   &$^{127}_{53}{I}$&  545.40& 4262.30&1253.00      \\ \vspace{2mm}
$^{40}_{18}{Ar}$& 166.50 &560.49&228.00   &$^{139}_{57}{La}$& 581.50& 4787.43&1400.00      \\ \vspace{2mm}
$^{39}_{19}{K}$&  78.80  &285.14&123.55   &$^{181}_{73}{Ta}$& 907.30& 7912.80&2358.00      \\ \vspace{2mm}
$^{40}_{20}{Ca}$& 38.00  &135.32&61.25    &$^{208}_{82}{Pb}$& 902.10& 7857.37&2643.00      \\ \vspace{2mm}
$^{51}_{23}{V}$&  198.40 &879.18&323.20   &$^{209}_{83}{Bi}$& 824.38& 8257.85&2497.00      \\ 
 \end{tabular}
\end{ruledtabular}
\end{table}
\begin{table}[h]
\caption{$<\sigma>(10^{-42}{cm}^2)$ for the inclusive reaction for some nuclei. }
\begin{ruledtabular}
\begin{tabular}{cccc}
Nucleus & Experimental& Theoretical & Present\\
 & Results& Results& Calculation.\\ \hline\\
&&14.00\cite{kosmas}, 14.00\cite{nieves}&13.60\\
&&16.40\cite{volpe}, 12.30\cite{hayes}&\\
$^{12}{C}$&$13.2\pm 0.5\pm 1.3$\cite{expcar1}, $14.8\pm 0.7 \pm 1.4$\cite{expcar3}& 14.40\cite{kolbe}, 12.90, 17.60\cite{auer}&\\\hline\\
$^{16}{O}$&&16.90, 17.20\cite{auer}&14.50\\ \hline\\
$^{56}{Fe}$&$256\pm 108\pm 43$\cite{expfe}&240\cite{kolbe2}, 214\cite{mintz1}&277\\ \hline\\
$^{208}{Pb}$&&3620\cite{kolbe2},  2954, 3204\cite{suzuki}, 4439\cite{volpe1}&2643\\
\end{tabular}
\end{ruledtabular}
\end{table}

\end{document}